\documentclass[prl,aps,showpacs,twocolumn]{revtex4}
\usepackage{epsfig}
\usepackage{amsmath}

\begin{document}

\title{Berry Phase in Atom-Molecule Conversion Systems and Fractional
Monopole }
\author{Li-Bin Fu$^{1,2}$ and Jie Liu$^{1,2}$}
\affiliation{1. Center for Applied Physics and Technology, Peking University, Beijing
100084, China \\
2. Institute of Applied Physics and Computational Mathematics, P.O. Box
8009, Beijing 100088, China}

\begin{abstract}
We investigate the geometric phase or Berry phase of adiabatic
quantum evolution in an atom-molecule conversion system, and find
that the Berry phase in such system  consists of two parts: the
usual Berry connection term  and a novel term from the nonlinearity
brought forth by the atom-molecule conversion. The geometric phase
can be viewed as the flux of the magnetic field of a monopole
through the surface enclosed by a closed path in parameter space.
The charge of the monopole, however, is found to be one third of the
elementary charge of the usual quantized monopole.
\end{abstract}

\maketitle

Berry phase \cite{berry}, which reveals the gauge structure
associated with a phase shift in adiabatic processes in quantum
mechanics, has attracted great interest in physics \cite{liu2}. One
classical example of Berry phase is a spin half particle aligned to
a magnetic field, and the field is made to rotate adiabatically in a
3-D parameter space (see Fig.1 a). The Berry phase of such system
has been interpreted as the flux of a magnetic field of a quantized
monopole through the surface enclosed by the loop in parameter
space.

On the other hand, association of ultracold atoms into molecules is
currently a topic of much experimental and theoretical interest
\cite{rev} with important applications ranging from the search for
the permanent electric dipole moment \cite{bin1} to BCS-BEC
(Bose-Einstein condensate) crossover physics \cite{bin2}. Through
Feshbach resonance \cite{fesh} or photoassociation \cite{photo}, a
pair of atoms can convert into a bounded molecule. The atom-molecule
conversion under mean field treatment is governed by a nonlinear
Schr\"{o}dinger equation, in which the nonlinearity is from the fact
that two or more atoms are needed to form one molecule\cite{meanf}.
Since the adiabatic manipulation is an optimal way to yield high
conversion efficiency, great efforts and big progress have been made
towards the adiabatic condition\cite{hanpu} and
adiabaticity\cite{itin} of the nonlinear quantum evolution.
Nevertheless, the knowledge of the Berry phase for the adiabatic
evolution in such nonlinear system is very limited. This system not
only lacks superposition principle due to the presence of
nonlinearity\cite{kavsh} but also have no $U(1)$-invariance because
the chemical potentials of atomic component and molecular component
are not identical\cite{meng}.
%%%%%%%%%%%%%%%%%%%%%%%%%%%%%%%%%%%%%%%%%%%%%%%%%%%%%%%%%%%%%%%%%%%%%%%%%%%%%%%%%%%%%%%%%
\begin{figure}[b]
\begin{center}
\rotatebox{0} {\resizebox *{7.5cm}{7.5cm} {\includegraphics[bb=0 0
553 609] {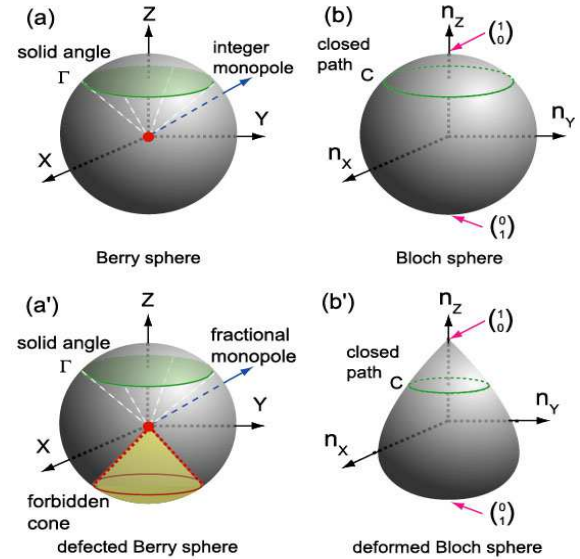}}}
\end{center}
\caption{(color online) The parameter space (a) and Bloch sphere (b)
for a spin half particle in magnetic field. (a') and (b') are the
parameter space and Bloch sphere for the atom-molecule conversion
system, respectively. The parameters change adiabatically along a
close path shown as the green circles in the parameter spaces or
Berry spheres . Accordingly, the eigenstate will evolve and form a
close path schematically plotted as green circles on the Bloch
spheres.
The gray cone in (a') is the boundary for which $\protect\theta =2%
\protect\pi /3. $, inside which, i.e., $\protect\theta >2\protect\pi
/3$, no eigenstate exists. See text for details.} \label{fig1}
\end{figure}
%%%%%%%%%%%%%%%%%%%%%%%%%%%%%%%%%%%%%%%%%%%%%%%%%%%%%%%%%%%%%%%%%%%%%%%%%%%%%%%%%%%%%%%%%%

In this letter we formulate the adiabatic geometric phase in a
general formulism for the atom-molecule conversion systems and
derive the explicit expression of the Berry phase analytically. We
find strikingly that  the circuit integral of Berry connection of
the instantaneous eigenstate alone can not account for the geometric
phase, while a novel term due to  the nonlinearity brought forth by
the atom-molecule coupling emerges. Only with the inclusion of this
additional contribution, the total geometric phase can be
interpreted as a flux of a magnetic field of a monopole through the
surface enclosed by the closed path in parameter space. There exists
a forbidden cone in Berry sphere (i.e., parameter space, see Fig. 1
a') and the Bloch sphere representing projective Hilbert space is
deformed dramatically (Fig.1 b'). The charge of the monopole is
found to be one third of the elementary charge of the usual
quantized monopole.

Let us consider an atom-molecule system with energy $\mathcal{H}(\hat{\psi},%
\hat{\psi}^{\dagger };\mathbf{R}),$in which $\mathbf{R}$ denotes all the
system parameters that vary in time slowly, $\hat{\psi}=\{\hat{\psi}_{i}\}$
and $\hat{\psi}^{\dagger }=\{\hat{\psi}_{i}^{\dagger }\}$ (with $i=1,\cdots,
N)$ are the annihilation and creation field operators for atom with $%
i=1,\cdots ,M$ and for molecule with $i=M+1,\cdots ,N.$ They obey the
commutation relations $\left[ \hat{\psi}_{i},\hat{\psi}_{i}^{\dagger }\right]
=\delta _{ij}$ for bosons. Under the mean field treatment, $\hat{\psi}$ and $%
\hat{\psi}^{\dagger }$ are replaced by complex numbers $\psi $ and $\psi
^{\ast }$. It is convenient to write $\mathcal{H}(\psi ,\psi ^{\ast };%
\mathbf{R})=\sum_{i,j}\psi _{i}^{\ast }T_{ij}(\psi ,\psi ^{\ast };\mathbf{R}%
)\psi _{j},$ then with the help of the above commutation relations, we
obtain following nonlinear Schr\"{o}dinger equations ($\hbar =1$)
\begin{equation}
i\frac{d\psi _{j}}{dt}=\sum_{k}H_{jk}(\psi ,\psi ^{\ast };\mathbf{R})\psi
_{k},  \label{sc}
\end{equation}
in which the Hamiltonian
\begin{equation}
H_{jk}(\psi ,\psi ^{\ast };\mathbf{R})=T_{jk}(\psi ,\psi ^{\ast };\mathbf{R}%
)+\sum_{i}\psi _{i}^{\ast }\frac{\partial T_{ik}}{\partial \psi _{j}^{\ast }}%
.  \label{ham}
\end{equation}

The above system is not invariant under usual $U(1)$ transformation,
instead, it is invariant under the following co-diagonal $U(1)$
transformation,
\begin{equation}
U(\eta )=e^{i\Theta (\eta )},\Theta (\eta )=\left(
\begin{array}{cc}
\eta I^{M} & 0 \\
0 & \kappa \eta I^{N-M}%
\end{array}
\right) ,  \label{co}
\end{equation}
in which $I^{K}$ is the $K$ rank unit matrix and $\kappa $ is an integer
related with molecule structure, e.g. for diatomic molecule $\kappa =2$ \cite%
{meng}$.$ Obviously, for $\kappa =1$ the system reduces to an atomic system
and $U(\eta )$ is just the ordinary $U(1)$ transformation.

The eigenequation of the above system is
\begin{equation}
\sum_{k}H_{jk}(\overline{\phi }(\mathbf{R}),\overline{\phi }^{\ast }(\mathbf{%
R});\mathbf{R})\overline{\phi }_{k}(\mathbf{R})=\mu _{j}(\mathbf{R})%
\overline{\phi }_{j}(\mathbf{R}),  \label{eig}
\end{equation}
where $\mu _{j}(\mathbf{R})=\mu (\mathbf{R})$ for $j\leq M$ (atom) and $\mu _{j}(%
\mathbf{R})=\kappa \mu (\mathbf{R})$ for $j>M$(molecule).

The above eigenequation defines the eigenfunction $\bar\phi$ and the
eigenvalue (or chemical potential) $\mu$ that are functions of the adiabatic
parameter $\mathbf{R}$.

The equation (\ref{sc}) and its conjugate construct a canonical structure of
classical dynamics with the energy $\mathcal{H}(\psi ,\psi ^{\ast };\mathbf{R%
})$ as classical Hamiltonian and $\left( \psi ,i\psi ^{\ast }\right)
$ as a canonical variable pair. The gauge symmetry of $\mathcal{H}$
given by (\ref{co}) implies that the total atoms number is conserved
and the dynamics of the overall phase can be separated from the rest
of the degrees of freedom\cite{wuliu03}. For simplicity and without
losing generality we denote$\lambda
=\arg \psi _{1}$ and set total phase as $\lambda$ for atomic components and $%
\kappa\lambda$ for molecular components, respectively. The
normalization condition is $\sum_{i=1}^{M}|\psi _{i}|^{2}+\kappa
\sum_{j=M+1}^{N}|\psi _{j}|^{2}=1.$ The other variables form a close
set of Hamiltonian dynamics. We choose a new set of canonical
variables $(q,p)$ as $q=\left( q_{1},\cdots ,q_{i},\cdots
,q_{N-1}\right) $ with $q_{i}=-\arg (\psi _{i+1})+\kappa _{i+1}\arg
(\psi _{1})$, $p=\left( p_{1},\cdots ,p_{i},\cdots ,p_{N-1}\right) $
with $p_{i}=|\psi _{i+1}|^{2}$, in which $\kappa _{i}=1$ for $i\leq
M$ and $\kappa _{i}=\kappa $ for $i>M.$ From (\ref{sc}) and its
canonical structure and using the normalization condition, we obtain
the dynamical equations for the overall phase and other variables,
\begin{eqnarray}
\frac{d\lambda }{dt} &=&p\frac{dq}{dt}-\mathcal{H}(p,q)-\Lambda (p,q),
\label{pas} \\
\dot{p} &=&-\frac{\partial \mathcal{H}}{\partial q},\dot{q}=\frac{\partial
\mathcal{H}}{\partial p},  \label{ca}
\end{eqnarray}
in which $\Lambda (p,q)=Re\left( \sum_{i,j,k}\psi _{j}^{\ast }\psi
_{i}^{\ast }\frac{\partial T_{ik}}{\partial \psi _{j}^{\ast }}\psi
_{k}\right) =Re\left( \sum_{i,j,k}\sqrt{p_{i}}\left( p_{j}\frac{\partial
\widetilde{T}_{ik}}{\partial p_{j}}-i\frac{\partial \widetilde{T}_{ik}}{%
\partial q_{j}}\right) \sqrt{p_{k}}\right) $ and $\widetilde{T}%
_{ik}(p,q)=e^{i(\arg (\psi _{k})-\arg (\psi _{i}))}T_{ik}.$

For a linear quantum case, both $H_{ij}$ and $T_{ij}$ are the
functions of the parameter $\mathbf{R}$ only, so that the last term
in Eq.(5) vanishes, i.e., $\Lambda (p,q)=0$. The second term in the
right-hand of Eq.(5) is the energy, whose time integral gives so
called dynamical phase. The time integral of the first term is the
Aharonov-Anandan phase for a cyclic quantum evolution\cite{aa}. The
above observation is readily extended to the adiabatic evolution of
a quantum eigenstate, because the adiabatic theorem of quantum
mechanics dictates that an initial nondegenerate eigenstate remains
to be an instantaneous eigenstate and the evolution will be cyclic
when the parameters move slowly in a circuit. In this case, the
second term is the eigenenergy and the first term is just the Berry
connection, i.e. $i<\bar{\phi}(\mathbf{R})|\nabla
|\bar{\phi}(\mathbf{R})>$. Then the Berry phase equals to the
circuit integral of the Berry connection.

However, for our nonlinear system, the contribution of the last term in (\ref%
{pas}) should be taken into account. Notice that the chemical potential is
usually not identical to the energy while the dynamic phase should be the
time integral of the chemical potential, we need to evaluate following
quantity in adiabatic limit,
\begin{equation}
\Xi (p,q;\mathbf{R})=\mathcal{H}(p,q)+\Lambda (p,q)-\mu (\mathbf{R}),
\label{new}
\end{equation}
We denote $p=\overline{p} (\mathbf{R})+\delta p$ and $q=\overline{q}(\mathbf{%
R})+\delta q$ . Here $\overline{p}(\mathbf{R})$ and $\overline{q}(\mathbf{R}%
) $ are the fixed points of Eq.(6) that are local energy minima of system
and therefore correspond to the eigenstates defined by (4). The vector $%
(\delta p,\delta q)$ represents the correction to the adiabatic
eigenstates in the order of
$|\frac{d\textbf{R}}{dt}|$\cite{excitation}. As will be shown,
$(\delta p,\delta q)$ contains some secular terms in addition to the
rapid oscillations. These secular terms will be accumulated in the
nonlinear adiabatic evolution and contribute to the geometric phase.

We expand the quantity $\Xi (p,q;\mathbf{R})$ around the fixed point,
\begin{equation}
\Xi (p,q;\mathbf{R})=\frac{\partial \Lambda} {\partial p}|_{(\bar p,\bar
q)}\delta p +\frac{\partial \Lambda} {\partial q}|_{(\bar p,\bar q)}\delta q
+ o(\delta q^{2},\delta p^{2}).
\end{equation}
Here we use the relations: $\mathcal{H}(\bar p,\bar q)+\Lambda (\bar p,\bar
q)=\mu (\mathbf{R})$ and $\partial \mathcal{H}(p, q)/\partial p |_{\bar p,
\bar q} = \partial \mathcal{H}(p, q)/\partial q|_{\bar p, \bar q} =0$

On the other hand, the $(\delta p,\delta q)$ can be evaluated from
following Hamiltonian equations,
\begin{eqnarray}
\dot{q} &=&\frac{\partial ^{2}\mathcal{H}}{\partial p\partial p}|_{(\bar{p},%
\bar{q})}\delta p+\frac{\partial ^{2}\mathcal{H}}{\partial p\partial q}|_{(%
\bar{p},\bar{q})}\delta q+o(\delta q^{2},\delta p^{2}), \\
\dot{p} &=&-\frac{\partial ^{2}\mathcal{H}}{\partial q\partial p}|_{(\bar{p},%
\bar{q})}\delta p-\frac{\partial ^{2}\mathcal{H}}{\partial q\partial q}|_{(%
\bar{p},\bar{q})}\delta q+o(\delta q^{2},\delta p^{2}).
\end{eqnarray}
Omitting the higher order terms, keeping the secular terms by average over
the fast oscillations, we obtain
\begin{equation}
(\left\langle \delta p\right\rangle ,\left\langle \delta q\right\rangle
)^{T}=\Omega ^{-1}(\frac{d\overline{p}}{d\mathbf{R}}\mathbf{\dot{R}},\frac{d%
\overline{q}}{d\mathbf{R}}\mathbf{\dot{R}})^{T},
\end{equation}
where the matrix $\Omega =\left(
\begin{array}{cc}
-\frac{\partial ^{2}\mathcal{H}(p,q)}{\partial q\partial p} & -\frac{%
\partial ^{2}\mathcal{H}(p,q)}{\partial q \partial q} \\
\frac{\partial ^{2}\mathcal{H}(p,q)}{\partial p\partial p} & \frac{\partial
^{2}\mathcal{H}(p,q)}{\partial p \partial q}%
\end{array}
\right)|_{(\bar{p},\bar{q})} $ is the Hessian matrix of the classical
Hamiltonian.

Combining Eq.(8) and (11) and with the help of Eq.(5), we find that,
except for the dynamical phase (i.e., the time integral of the
chemical potential), the eigenstate of the nonlinear atom-molecule
conversion system acquires following additional phase during the
adiabatic cyclic evolution\cite{note},
\begin{eqnarray}
\gamma _{g} &=&\oint \overline{p}\frac{d\overline{q}}{d\mathbf{R}}\cdot d%
\mathbf{R}  \notag \\
&-&\oint (\frac{\partial \Lambda }{\partial p},\frac{\partial \Lambda }{%
\partial q})|_{(\bar{p},\bar{q})}\cdot \Omega ^{-1}\cdot (\frac{d\overline{p}%
}{d\mathbf{R}},\frac{d\overline{q}}{d\mathbf{R}})^{T}\cdot d\mathbf{R}.
\label{ggg}
\end{eqnarray}%
In contrast to previous works \cite{biao}, the adiabatic geometric
phase in the atom-molecule system is dramatically modified. The
first term is the usual expression of the Berry phase that can be
rewritten as  the circuit integral of the Berry connection $\frac{i%
}{2}\left( \left\langle \bar{\phi}(\mathbf{R})\right\vert \left. \nabla \bar{%
\phi}(\mathbf{R})\right\rangle -\left\langle \nabla \bar{\phi}(\mathbf{R}%
)\right\vert \left. \bar{\phi}(\mathbf{R})\right\rangle \right) $.
The novel  second term indicates that, the high-order correction to
adiabatic approximate solution that is negligible in linear case,
could be accumulated in the nonlinear adiabatic evolution with an
infinite time duration in adiabatic limit and contributes a finite
phase with geometric nature.

As an illustration, we consider following atom-molecule conversion model
whose energy takes the form,
\begin{eqnarray}
\mathcal{H} &=&\frac{R\cos \theta }{2}\left( \hat{\psi}_{1}^{\dagger }\hat{%
\psi}_{1}-\hat{\psi}_{2}^{\dagger }\hat{\psi}_{2}\right) +  \notag \\
&&\sqrt{\frac{3}{8}}\frac{R\sin \theta }{2}\left( e^{-i\phi }\hat{\psi}%
_{1}^{\dagger }\hat{\psi}_{1}^{\dagger }\hat{\psi}_{2}+h.c.\right) ,
\label{en}
\end{eqnarray}
where $\hat{\psi}=(\hat{\psi}_{1},\hat{\psi}_{2})$ and $\hat{\psi}^{\dagger
}=\{\hat{\psi}_{1}^{\dagger },\hat{\psi}_{2}^{\dagger }\}$ are the
annihilation and creation operators for atom and molecule respectively, the
terms $\hat{\psi}_{1}^{\dagger }\hat{\psi}_{1}^{\dagger }\hat{\psi}_{2}+h.c.$
describe coupling between atom pairs and diatomic molecules, and $\mathbf{R}%
=(R,\theta ,\varphi )$ are parameters. Obviously, the system is invariant
under the transformation $U(\eta )=e^{i\Theta (\eta )},\Theta (\eta )=\left(
\begin{array}{cc}
\eta & 0 \\
0 & 2\eta%
\end{array}
\right) $.

Let us rewrite $\mathcal{H}(\psi ,\psi ^{\ast };\mathbf{R})=\sum_{i,j}\psi
_{i}^{\ast }T_{ij}(\psi ,\psi ^{\ast };\mathbf{R})\psi _{j},$ where the
matrix elements $T_{11}=-T_{22}=\frac{R\cos \theta }{2},$ $%
T_{12}=T_{21}^{\dagger }=\sqrt{\frac{3}{8}}\frac{R\sin \theta }{2}%
e^{-i\varphi }\hat{\psi}_{1}^{\dagger }$ and the nonlinear
Schr\"{o}dinger equation takes the form of  Eq.(1) with
\begin{equation}
H(\psi ,\psi ^{\ast };\mathbf{R})=\left(
\begin{array}{cc}
\frac{R\cos \theta }{2} & \sqrt{\frac{3}{8}}e^{-i\varphi }R\sin \theta \psi
_{1}^{\ast } \\
\sqrt{\frac{3}{8}}e^{i\varphi }R\sin \theta \psi _{1}/2 & -\frac{R\cos
\theta }{2}%
\end{array}
\right) ,  \label{hami}
\end{equation}
where $\psi _{1}$ and $\psi _{2}$ are complex amplitudes for atom and
molecule respectively. The projective Hilbert space is spanned by the vector
$\mathbf{n}_{a}=(2\sqrt{2}\ \emph{Re}[(\psi _{1}^{\ast })^{2}\psi _{2}],2%
\sqrt{2}\emph{Im}((\psi _{1}^{\ast })^{2}\psi _{2}),|\psi _{1}|^{2}-2|\psi
_{2}|^{2}).$ Obviously, every point in this space corresponds to a class of
quantum states among which the states are only different in co-diagonal
total phases (see Eq. (\ref{co})). With the normalization condition $|%
\overline{\psi }_{1}|^{2}+2|\overline{\psi }_{2}|^{2}=1,$ we plot the
projection space in Fig. 2(b'), which is a ''tear-drop'\ shaped surface\cite%
{vardi}.

The eigenequations take the forms of $H(\overline{\phi },\overline{\phi }%
^{\ast };\mathbf{R})\left(
\begin{array}{c}
\overline{\phi }_{1} \\
\overline{\phi }_{2}%
\end{array}
\right) =\left(
\begin{array}{cc}
\mu & 0 \\
0 & 2\mu%
\end{array}
\right) \left(
\begin{array}{c}
\overline{\phi }_{1} \\
\overline{\phi }_{2}%
\end{array}
\right) $. The eigenequations are solved and the eigenfunctions are obtained
as follows,
\begin{equation}
\overline{\phi }_{2}^{\pm }=\frac{\left( -\cos \theta \pm 1\right) }{\sqrt{6}%
\sin \theta },\overline{\phi }_{1}^{\pm }=e^{i\varphi /2}\sqrt{1-2|\overline{%
\phi }_{2}^{\pm }|^{2}},  \label{dda}
\end{equation}
with the eigenvalue (or chemical potential) $\mu _{\pm }=\frac{R}{4}\left(
\cos \theta \pm 1\right) $.

Following our general formalism, we choose the total phase as $\lambda =\arg
\psi _{1}$ and define the canonical pair as $q=-\arg \psi _{2}+2\arg \psi
_{1}$ and $p=|\psi _{2}|^{2}$. Hence, we have $\overline{p}=|\overline{\phi }%
_{2}^{\pm }|^{2}$ and $\overline{q}=\varphi .$ Substituting the concrete
expressions of matrix $\{T_{ij}\}$ to the definition of the quantity $%
\Lambda $, we get $\Lambda (p,q)=\sqrt{\frac{3}{8}}\frac{R\sin \theta }{2}%
(1-2p)\sqrt{p}\cos \left( q-\varphi \right) $. On the other hand, \ from the
system energy $\mathcal{H}(p,q)=\frac{R\cos \theta }{2}\left( 1-3p\right) +%
\sqrt{\frac{3}{8}}R\sin \theta (1-2p)\sqrt{p}\cos (q-\varphi ),$ we get $%
\Omega ^{-1}=\frac{\sqrt{8}}{\sqrt{3}R\sin \theta }\left(
\begin{array}{cc}
0 & -\frac{2\bar{p}\sqrt{\bar{p}}}{(1+6\bar{p})} \\
\frac{1}{(1-2\bar{p})\sqrt{\bar{p}}} & 0%
\end{array}%
\right) .$ \ After lengthy calculation, we obtain the Berry phase according
to formula (\ref{ggg}),
\begin{eqnarray}
\gamma _{g} &=&\oint \overline{p}d\varphi +\oint \frac{(1-6\overline{p})%
\overline{p}}{1+6\overline{p}}d\varphi  \label{gggg} \\
&=&\frac{1}{6}\oint \left( 1\mp \cos \theta \right) d\varphi ,  \label{g2}
\end{eqnarray}%

The above theoretical formulation on the Berry phase has been
verified numerically by directly integrating the Schr\"{o}dinger
equation. On the other hand, we recognize that the above system
although admits the quantal equations of motion, appears formally to
have classical structure if we regard the total phase and total
particle number as a pair of canonical conjugate variables. We thus
could exploit this particular feature to construct a canonical
transformation to action-angle variables. With including the
canonical motions in projective Hilbert space represented by Eq.(6),
we have derived the Hannay's angle of a geometric nature associated
with the adiabatic evolution. The Hannay's angle is found to exactly
equal to minus Berry phase of Eq.(\ref{gggg})\cite{preparation}.
This fact indicates a novel connection between Berry phase and
Hannay's angle in contrast to the usual derivative
form\cite{hannay}, and supports our new expression of Berry phase
from the other aspect.

For the linear systems, such as the spin-half system, i.e.,
$H=-\frac{1}{2}\mathbf{R\cdot \sigma }$ where $\mathbf{\sigma }$ are
pauli matrices and $\mathbf{R=(}R\sin \theta \cos \phi ,R\sin \theta
\sin \phi ,R\cos \theta \mathbf{)}$ is a vector in the 3-D parameter
space. The Berry phase equals to the circuit integral of the Berry
connection and is interpreted as the flux of the magnetic field of a
virtual quantized monopole through the surface enclosed by the loop
in parameter space, i.e., $B_{m}=g_{0}\frac{\mathbf{R}}{R^{3}}$ with
the
elementary charge $g_{0}=\frac{1}{2%
}$. In general, the degeneracies of the spectrum in parameter space
play an important role in connexion with the geometric phase. Each
degeneracy can be seen as a charge distribution located at the
contact point between energy surfaces. Because the eigenstates are
smooth and single valued outside the degeneracies, the total charge
of the distribution, i.e., the monopole charge is necessarily an
integer multiple of the elementary charge $g_{0}=1/2 $. The
non-elementary monopoles with integer multiples of $g_{0}$  have
been found in case of light propagating and in condensed matter
physics \cite{berry86,simon}. The mechanism for the production of
monopole charges larger than the elementary $g_{0}$ is due to
constraints that act on the system\cite{leb}.

For our nonlinear system, when the parameters $\mathbf{R}=(R,\theta
,\varphi )$ are considered as spherical coordinates of a vector in a
3-D space, and then from (\ref{g2}) we get the vector potential,
$
\mathbf{A}=\frac{1}{6}\frac{\left( 1-\cos \theta \right) }{R\sin \theta }%
\widehat{e}_{\varphi }$. Here, for convenience we only consider the
branch of $\mu _{+}$. Hence, the Berry phase of atom-molecule
conversion system can also be interpreted as the flux of a magnetic
field of a virtual monopole through the surface
enclosed by the closed path in parameter space (see in fig. 2(a')),
that is,  $\mathbf{%
B}=\bigtriangledown \times \mathbf{A}_{spin}=g\frac{\mathbf{R}}{R^{3}}%
\mathbf{.}$ Strikingly, the monopole charge $g=\frac{1}{3}g_{0}$,
one third of the elementary charge. We attribute the fractional
charge to the symmetry
breaking of the parameter space by the boundary. From (\ref{dda}), we see $|%
\overline{\psi }_{2}|$ increases with $\theta $ monotonously. When $\theta
=2\pi /3,$ $|\overline{\psi }_{2}|=1/\sqrt{2}$ reaching its extreme value
(since $|\overline{\psi }_{1}|^{2}+2|\overline{\psi }_{2}|^{2}=1).$ It
implies that there is no eigenstate in the regime $\theta >2\pi /3,$ i.e.,
the Berry sphere of this system is a defected sphere with a forbidden cone
bounded by $\theta =2\pi /3.$
This curious structure has been illustrated in
fig. 1(a').

In summary, we have investigated the adiabatic geometric phase in
the atom-molecule conversion systems. A novel formula of geometric
phase is derived and an exotic monopole with fractional elementary
charge is found. The above phase and monopole are expected to be
observed in the future ultracold atom experiments.

This work is supported by National Natural Science Foundation of
China (No.10725521,10604009), 973 project of China under Grant No.
2006CB921400, 2007CB814800.


\begin{thebibliography}{99}
\bibitem{note} This is for the atom components. For the molecule
components, the phase should time the factor $\kappa$.
\bibitem{preparation} Detailed calculation on the Hannay's angle  will be presented elsewhere.
\end{thebibliography}
\end{document}